\documentclass[12pt]{article}
\usepackage{graphicx}
\usepackage{subfigure}
\usepackage{amsmath}
\usepackage{amssymb}

\begin{document}

\begin{center}
\begin{large}
\title\\{\textbf{Masses of Heavy Flavour Mesons in a potential Model Approach
with Wave Function containing Airy's Infinite Series.}}
\end{large}

\addvspace{0.2in}

\author\

\title\\{\textbf{$Abdul\;Aziz^{\emph{a,b}}\footnotemark,Sabyasachi\;Roy^{\emph{a}}\;\;and\:Atri\:Deshamukhya^{\emph{b}}$}} \\\

\footnotetext{Corresponding author. e-mail:  \emph{aaziz.phys@gmail.com}}
\textbf{a}. Department of Physics, Karimganj College, Karimganj, Assam-788710.\\
\textbf{b}. Department of Physics, Assam University, Silchar, Cachar-788001. \\

\begin{abstract}
We report the masses of $B$ and $D$ sectors heavy-flavoured mesons obtained by using our recently developed meson wave function employing potential model approach with linear confinement term in potential as parent in the perturbation method. As the wave-function involves infinite Airy's polynomial series, in carrying out the mass calculation, to avoid divergences, we have introduced some cut-off parameter for inter-quark separation. Our results for ground state masses of heavy-flavoured $B$ and $D$ sector mesons are reasonably closer to the PDG masses.

\end{abstract}
\end{center}
\

\section{Introduction:}\rm
Studies on static and dynamic properties of heavy-flavoured mesons following non-perturbatively  is a reliable approach to throw light on QCD at low energy which is not directly explorable otherwise. In that, the studies on mass spectrum is very important. For a meson, which is a two-quark bound system, the inter quark potential is mainly generated by gluon dynamics, under static approximation, with valence quark effect. At distances smaller than 1 Fermi, the quark-antiquark potential is Coulombic, due to asymptotic freedom. At large distances the potential should be linear due to formation of confining flux tubes. This allows the simplest and generally accepted Cornell potential for the studies of the quark dynamics inside a meson. \\
The studies of properties of mesons in the low energy regime with such linear plus Coulombic potential is challenging due to its non-perturbative nature. However, in recent past phenomenological studies in potential model approach has resulted in reliable outputs. The wave function for the heavy-light mesons have been calculated earlier within the framework of QCD potential model \cite{DKC1, Eichten, KKP} with considerable accuracy \cite{DKC2}. This has been deduced both with Coulombic term in potential as parent\cite{DKC3} and also with the linear confinement term as parent\cite{DKC4}.\\
Mass, on the other hand, is a fundamental parameter - studies on which always gives new insight into the system. Recent works on masses of heavy-flavoured mesons at highr dimension have attracted considerable attention to the subject \cite{SRmass}. In this work we have done studies on masses of heavy-flavoured mesons in non-perturbative potential model approach using wave function developed in \cite{SR1}. The wave function used contains Airy's infinite series which is otherwise divergent with infinite limit of integration. Under this limitation, in this work we have followed some reasonable limit to integration guided by convergence condition of the wave function. The masses for B and D sector mesons are calculated using some standard relations which are then compared with the available PDG (Particle Data Group) masses \cite{PDG}. The results are found to be satisfactory within the limitation of the approach. \\
Section 2 contains brief review of formalism followed, section 3 contains the calculation of masses. In section 4 we report our results and section 5 contains conclusion and remarks.

\section{Formalism:}\rm

\subsection{Meson Mass:}
Pseudoscalar meson mass can be computed from the following relation \cite{KKP}
\begin{eqnarray}
M_P = m_Q +m_{\overline{Q}}+ \triangle E  \\
where,\;  \triangle E = <H> \nonumber
\end{eqnarray}
Hamiltonian operator $H$ has the form:
\begin{equation}
H=-\frac{\nabla^2}{2\mu}+V(r)
\end{equation}
Here,$\mu=\frac{m_Q m_{\overline{Q}}}{m_Q+ m_{\overline{Q}}}$ is the reduced mass of the meson with $m_Q$ and $m_{\overline{Q}}$ are the quark and antiquark masses; $V(r)$ is the inter-quark potential.
\begin{equation}
 \nabla^{2}\equiv \frac{d^{2}}{dr^{2}}+\frac{2}{r}\frac{d}{dr}-\frac{l(l+1)}{r^{2}}
\end{equation}
For ground state meson ( $l=0$),
\begin{equation}
 \nabla^{2}\equiv \frac{d^{2}}{dr^{2}}+\frac{2}{r}\frac{d}{dr}
\end{equation}

\subsection{Potential Model:}
The cornell potential signifying bound state quark-antiquark system is of the form :
\begin{equation}
V (r) = -C_F \frac{\alpha_s}{r} + br + c
\end{equation}
Here, $\alpha_s$ is the strong coupling constant, $b$ is the confinement parameter and $c$ is a constant added to the potential to scale it suitably.
$ C_F $ is the colour factor, which is given by :
\begin{equation}
C_F = \frac{N_C ^{2}-1}{2N_c}
\end{equation}
$ N_C $ is the colour quantum number; for $ N_C = 3 $, we have $ C_F = \frac{4}{3} $. \\
\begin{equation}
V (r) = - \frac{4\alpha_s}{3r} + br + c
\end{equation}

\subsection{Wave Function with linear term as parent:}

As, with such a $linear\; plus \;Coulombic$ type of potential, it is not possible to extract exact solution of Schrodinger equation, we obtain meson wave function from two-body Schrodinger equation by applying quantum mechanical perturbation technique. With linear confinement term ($br$) in potential as parent in perturbation method, the wave function comes out in terms of Airy's infinite series \cite{SR1}.

\begin{equation}
\Psi^{total}(r)=\frac{N^{\prime}}{r}[1+ {A_1(r) r+A_2(r) r^{2}+A_3(r) r^{3}+..... }]A_i[\varrho_1 r+\varrho_0]
\end{equation}

Here, we have taken $\varrho_1$ $=$ $(2\mu b)^{1/3}$ and $ \varrho(r) = \varrho_1 r + \varrho_0 $. $N^\prime $ is the normalization constant of total wave function which is having the dimension of $GeV^{1/2}$. \\
Considering relativistic effect on the wave function following Dirac modification \cite{DKC3}, the total relativistic wave function is expressed as:
\begin{equation}
\Psi_{rel}^{tot} (r) = \frac{N^{\prime}}{r}[1+ {A_1(r) r+A_2(r) r^{2}+A_3(r) r^{3}+..... }]A_i[\varrho_1 r+\varrho_0](\frac{r}{a_0})^{-\epsilon }
\end{equation}
with,
\begin{equation}
a_0 = \frac{3 }{4\mu \alpha_s}= \frac{1}{B\mu} \;\; and\;\;  \epsilon = 1-\sqrt{1-(\frac{4\alpha_s}{3})^2}=1-\sqrt{1-(B)^2}
\end{equation}
It is to be mentioned that, the Airy's infinite series as a function of $\varrho =\varrho_1 r +\varrho_0 $ can be expressed as \cite{Airy} :
\begin{eqnarray}
A_i[\varrho] = a_1[1+\frac{\varrho^3}{6}+\frac{\varrho^6}{180}+\frac{\varrho^9}{12960}+...]-
 b_1[\varrho +\frac{\varrho^4}{12}+\frac{\varrho^7}{504}+\frac{\varrho^{10}}{45360}+...]
\end{eqnarray}
\begin{flushright}
 with $a_1=\frac{1}{3^{2/3} \Gamma(2/3)}=0.3550281$ and $ b_1=\frac{1}{3^{1/3} \Gamma(1/3)} =0.2588194.$ \\
\end{flushright}

$ A_1(r) , A_2(r), A_3(r) $ etc are obtained as \cite{SR1}:
\begin{eqnarray}
A_1(r)=-\frac{2\mu B}{2 \varrho_1 k_1 + \varrho_1 ^{2} k_2} \\
A_2(r)=-\frac{2\mu W^{\prime}}{2+4\varrho_1 k_1 + \varrho_1^{2} k_2}\\
A_3(r)= -\frac{2 \mu E A_1 }{6+6 \varrho_1 k_1 + \varrho_1^{2}k_2}  \\
A_4(r)= -\frac{2 \mu E A_2 - 2\mu b A_1}{12+ 8\varrho_1 k_1 + \varrho_1^{2}k_2} \\
A_5(r)= -\frac{2 \mu E A_3 - 2\mu b A_2}{20+ 10\varrho_1 k_1 + \varrho_1^{2}k_2} \\ \nonumber
\cdots\cdots\cdots\cdots \cdots\cdots\cdots\cdots \\ \nonumber
\cdots\cdots\cdots\cdots \cdots\cdots\cdots\cdots \\ \nonumber
\end{eqnarray}
$E$ and $W^\prime$ are the unperturbed and perturbed eigen energies.  We take $ k_1(r)$ and $ k_2 (r)$ to be some function of r from the identity:
\begin{eqnarray}
A_i ^{\prime}[\varrho]=\frac{d A_i[\varrho]}{dr}=Z(\varrho)A_i[\varrho] = [\frac{k_1(r)}{r}]A_i[\varrho]                                        \\
A_i ^{\prime\prime}[\varrho]=\frac{d^2 A_i[\varrho]}{dr^2}= [Z^{2}(\varrho)+Z^{\prime}(\varrho)]A_i(\varrho)=[\frac{k_2 (r)}{r^{2}}]A_i[\varrho]
\end{eqnarray}
so that,
\begin{eqnarray}
Z(\varrho)=\frac{k_1(r)}{r} \\
Z^{2}(\varrho)+Z^{\prime}(\varrho)=\frac{k_2 (r)}{r^{2}}
\end{eqnarray}
Considering lowest Airy polynomial order we obtain \cite{SR1}:
\begin{eqnarray}
k_1(r)=1+\frac{k}{r}\\
k_2(r)=\frac{k^2}{r^2}\\
with  \;\;\; k=\frac{a_1-b_1 \varrho_0}{b_1 \varrho_1}
\end{eqnarray}

\section{Calculation of Mass Terms:}

In the wave function, $\Psi_{rel}^{tot} (r)$  ( eqn. 9 ), considering only up to the term $A_2(r)$ in the first infinite series, we obtain:

\begin{eqnarray}
\Psi_{rel}^{tot} (r) = \frac{N^{\prime}}{r}[1+ B_1 r^3 + B_2 r^4 +B_3 r^5 +B_4 r^6 ]A_i[\varrho](\frac{r}{a_0})^{-\epsilon } \\
or,\; \Psi_{rel}^{tot}(r)= N a_0^{\epsilon}[ r^{-1-\epsilon} + B_1 r^{2-\epsilon}+ B_2 r^{3-\epsilon}+ B_3 r^{4-\epsilon}+ B_4 r^{5-\epsilon} ]A_i[\varrho] \\
where, \nonumber \\
B_1 = -\frac{2 \mu B}{\varrho_1^2 k^2}  \\
B_2 = \frac{4 \mu B}{\varrho_1^3 k^3} - \frac{2 \mu W^\prime}{\varrho_1^2 k^2}\\
B_3 = \frac{4 \mu B}{\varrho_1^3 k^4} + \frac{2 \mu W^\prime}{\varrho_1^4 k^4}4\varrho_1 k  \\
B_4 = \frac{2 \mu W^\prime(2+4\varrho_1)}{\varrho_1^4 k^4}  \\
\end{eqnarray}

With this simplified form of wave function, we then proceed to calculate meson masses. Following the expression of potential as in equation (7), we obtain from equation (2) ( for convenience we write $\Psi(r)$ for $\Psi_{rel}^{tot} (r)$ .):
\begin{eqnarray}
<H> = <-\frac{\nabla^2}{2\mu}> + <-\frac{4 \alpha_s}{3r}> + <\sigma r > + <c> \nonumber \\
= <H_1> + <H_2> + <H_3> + <H_4>
\end{eqnarray}
Here,
\begin{eqnarray}
<H_1> = \int_0^{\infty} 4 \pi r^2 \Psi(r)[-\frac{\nabla^2}{2\mu}]\Psi(r)dr \\
= \int_0^{\infty} 4 \pi r^2 \Psi(r)[-\frac{1}{2\mu}(\frac{d^2 \Psi(r)}{dr^2}+ \frac{2}{r}\frac{d \Psi(r)}{dr})]dr \\
<H_2> = \int_0^{\infty} 4 \pi r^2 (-\frac{B}{r})|\Psi(r)|^2 dr = - 4\pi B \int_0^{\infty} r |\Psi(r)|^2 dr  \\
<H_3> =  \int_0^{\infty} 4 \pi r^2 (b r)|\Psi(r)|^2 dr = 4\pi b \int_0^{\infty} r^3 |\Psi(r)|^2 dr  \\
<H_4> =  \int_0^{\infty} 4 \pi r^2 (c)|\Psi(r)|^2 dr = 4\pi c \int_0^{\infty}r^2 |\Psi(r)|^2 dr
\end{eqnarray}

$\frac{d\Psi(r)}{dr}$ and $ \frac{d^2\Psi(r)}{dr^2}$ involved in $<H_1>$ are calculated as :
\begin{eqnarray}
\frac{d\Psi(r)}{dr}= N a_0^{\epsilon}([(-1-\epsilon)r^{-2-\epsilon}+B_1(2-\epsilon)r^{1-\epsilon}+B_2(3-\epsilon)r^{2-\epsilon}+B_3(4-\epsilon)r^{3-\epsilon}+B_4(5-\epsilon)r^{4-\epsilon} ]Ai[\varrho]  \nonumber \\
+ [r^{-1-\epsilon}+ B_1 r^{2-\epsilon} + B_2 r^{3-\epsilon}+ B_3 r^{4-\epsilon}+ B_4 r^{5-\epsilon}]Ai^{\prime}[\varrho] )  \\
\frac{d^2\Psi(r)}{dr^2}= N a_0^{\epsilon}([(-1-\epsilon)(-2-\epsilon)r^{-3-\epsilon}+B_1(2-\epsilon)(1-\epsilon)r^{-\epsilon}+B_2(3-\epsilon)(2-\epsilon)r^{1-\epsilon} \nonumber \\
+B_3(4-\epsilon)(3-\epsilon)r^{2-\epsilon}+B_4(5-\epsilon)(4-\epsilon)r^{3-\epsilon}]Ai[\varrho] \nonumber \\
+2[(-1-\epsilon)r^{-2-\epsilon}+B_1(2-\epsilon)r^{1-\epsilon}+B_2(3-\epsilon)r^{2-\epsilon}+B_3(4-\epsilon)r^{3-\epsilon}+B_4(5-\epsilon)r^{4-\epsilon}]Ai^{\prime}[\varrho] \nonumber \\
+ [r^{-1-\epsilon}+ B_1 r^{2-\epsilon} + B_2 r^{3-\epsilon}+ B_3 r^{4-\epsilon}+ B_4 r^{5-\epsilon}]Ai^{\prime\prime}[\varrho] )
\end{eqnarray}

\section{Results:}
With the expressions for $ <H>$ we now proceed to compute pseudoscalar meson masses. Here,  the strong coupling constant $\alpha_s$ is take as 0.39 and  0.22 at charm and bottom quark mass scales, the value of the confinement parameter $b$  is 0.183 $GeV^2$ \cite{Charmonium} and the scale factor $c$ in the potential is taken as $1 GeV$ keeping conformity with our previous work \cite{NSB}, to make it compatible with meson masses in the calculations. The input values for quark masses are taken from ref \cite{Vinodkumar}, which are listed in Table-1. The masses for pseudoscalar $B$ and $D$ mesons  from PDG \cite{PDG}are shown in Table-2. Now, it is worth mentioning here that the wave function containing infinite Airy's polynomial series brings in divergences while carrying out the integrations with infinite upper limit as in the calculation of different terms involved in $<H>$. This inspires us to consider some reasonable cut-off to the upper infinite integration limit. In principle, this cut-off limit should be within the range of size of hadrons, i.e,  $ ~\frac{1}{M_p}$ ($M_p$ being the hadron mass). In our calculations we fix this cut-off out from the convergence condition of total wave function obtained through perturbation technique.
\begin{equation}
\Psi^{\prime}(r) < \Psi(r) \;\;i.e, \; |A_1(r) r + A_2(r) r^2 + \cdots \cdots | < 1
\end{equation}
This condition gives us the limiting values of cut-off parameter $r_0$ for different mesons. Considering upto the term $A_2(r)$ in the series, we obtain the cut-off limit from:
\begin{equation}
B_1 r^3 + B_2 r_0^4 + B_3 r_0^5 + B_4 r_0^6 =0
\end{equation}
The numerical values of this cut-off $r_0$ for different mesons are shown in Table-3. As, perturbed eigen energy $E$ and unperturbed energy $W^{\prime}$ are involved in the calculations of $r_0$, these are also reported in the same table. The Table-3 shows that the cut-off value should be greater than $3\;GeV^{-1}$. Consideration of this cut-off will not sacrifice the expected value or nature of the integrands; this is because Airy's function $Ai$ falls very sharply with $r$ beyond $r=3$ ($Ai[3]=0.0066 $)\cite{Airy}. \\
With that cut-off, then we calculate the energy expectation value $\triangle E$ for different mesons and compute the corresponding masses numerically. Results are shown in Table-4, in which we also record the ratio $\frac{\triangle E}{M_p}$ measured from our calculations. In Figure-1, we plot our masses along with standard PDG masses\cite{PDG}.

\section{Conclusion and remarks:}
In this work, we have developed formalism for the studies of meson masses with  QCD inspired potential model. Our results for ground state masses of heavy-light mesons are in close proximity with the standard PDG masses. The deviation on an average is less than $4$ percent from the standard masses (Table 4). That might be due to the truncation of the infinite Airy series in the calculation of masses. \\
Also, to avoid divergence arising due to infinite Airy series in calculations, we here introduce reasonable cut-off to infinite upper limit of integrations by invoking convergence condition of the Airy's series. That cut-off to $r_0$ is above $3.5$ $GeV^{-1}$ which is quite reasonable compared to the dimension of the mesons. \\
However, there is scope of refinement by considering higher terms of the Airy's series in the wave function. Also, the formalism may also be extended to measure masses of higher spectroscopic states of mesons. This work is in progress.


\begin{table}[tb]
\begin{center}
\caption{Quark masses from ref \cite{Vinodkumar}}
\begin{tabular}{|c||c|}
  \hline
    Quark &   Mass) \\
   \hline
  $m_{u/d}$     & 0.18 \; GeV    \\
  $m_s$         & 0.25 \; GeV    \\
  $m_c$         & 1.31 \;GeV    \\
  $m_b$         & 4.66 \;GeV    \\
 \hline
\end{tabular}
\end{center}
\end{table}

\begin{table}[tb]
\begin{center}
\caption{Meson masses from PDG}
\begin{tabular}{|c||c|}
  \hline
    Meson &   Mass(PDG) in GeV \\
   \hline
  $D^0 (c\overline{u})$         & 1.8649    \\
  $D_s (c\overline{s})$         & 1.9685    \\
  $B^0 (d\overline{b})$         & 5.2796    \\
  $B_s (s\overline{b})$         & 5.3668    \\
  \hline
\end{tabular}
\end{center}
\end{table}

\begin{table}[tb]
\begin{center}
\caption{Cut-off for different mesons}
\begin{tabular}{|c||c|c|c|}
  \hline
    Meson        & E (GeV) & $W^{\prime}$(GeV) & $r_0 (GeV^{-1})$ \\
   \hline
  $D^0 $         &  0.8428  & 0.4703   &  3.6052  \\
  $D_s $         &  0.7659  & 0.3867   &  3.5362  \\
  $B^0 $         &  0.8066  & 0.4264   &  3.5783  \\
  $B_s $         &  0.7215  & 0.3367   &  3.5021  \\
  \hline
\end{tabular}
\end{center}
\end{table}

\begin{table}[tb]
\begin{center}
\caption{Our calculated Meson masses (in GeV)}
\begin{tabular}{|c||c|c|c|c|}
  \hline
    Meson &                    $<H>=\triangle E$ & Our mass($M_p$) & PDG mass & $\triangle E /M_p $\\
   \hline
  $D^0 (c\overline{u})$       & 0.3603 & 1.8503 &  1.8649   & 0.1947 \\
  $D_s (c\overline{s})$       & 0.3997 & 1.9597 &  1.9685   & 0.2040 \\
  $B^0 (d\overline{b})$       & 0.4010 & 5.2410 &  5.2796   & 0.0765 \\
  $B_s (s\overline{b})$       & 0.4194 & 5.3294 &  5.3668   & 0.0787 \\
  \hline
\end{tabular}
\end{center}
\end{table}

\begin{figure}[h]
    \centering
    \includegraphics[width=7 in]{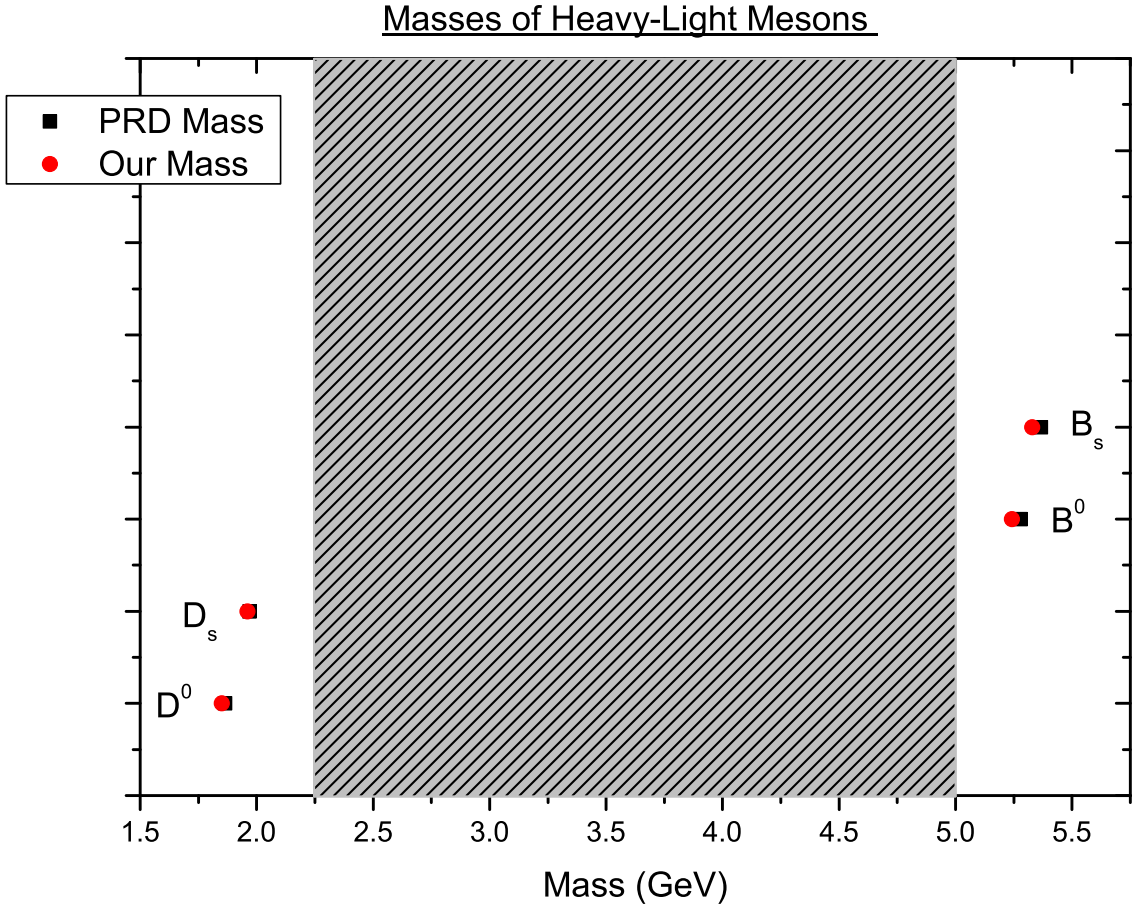}
    \caption{ Our Calculated Masses with PDG masses}
\end{figure}

\end{document}